\documentclass[12pt]{iopart}

\usepackage{color}
\usepackage{pifont}

\usepackage[normalem]{ulem}
\usepackage{graphicx,epstopdf}
\usepackage{textcomp}
\usepackage{tabularx}
\usepackage{booktabs}
\usepackage{array}
\usepackage{bbm}
\usepackage{bm}
\usepackage{iopams}
\usepackage{setstack}

\newtheorem{prop}{Proposition}\def\PRO{\begin{prop}}\def\ORP{\end{prop}}
\newtheorem{coro}{Corollary}\def\COR{\begin{coro}}\def\ROC{\end{coro}}
\newtheorem{theorem}{Theorem}\def\TH{\begin{theorem}}\def\HT{\end{theorem}}
\newtheorem{defin}[prop]{Definition}\def\DE{\begin{defin}}\def\ED{\end{defin}}
\newtheorem{lemme}[prop]{Lemma}\def\LE{\begin{lemme}}\def\EL{\end{lemme}}
\newtheorem{conjecture}{Conjecture}\def\TH{\begin{conjecture}}\def\HT{\end{conjecture}}

\providecommand{\ket}[1]{{|}#1\rangle}
\providecommand{\bra}[1]{\langle#1 {|}}
\providecommand{\kebra}[2]{\ket{#1}\bra{#2}}
\providecommand{\sprod}[2]{\langle#1|#2\rangle}

\providecommand{\idop}{\mathbbm 1}

\definecolor{dark_purple}{RGB}{152, 0, 70}

\begin{document}

\title{Adiabatic graph-state quantum computation}

\author{B Antonio$^1$, D Markham$^2$ and J Anders$^{1,3}$}
\ead{robert.antonio.10@ucl.ac.uk}

\address{$^1$Department of Physics and Astronomy, University College London, Gower Street, London WC1E 6BT, United Kingdom,}

\address{$^2$ CNRS LTCI, D\'{e}partement Informatique et R\'{e}seaux, Telecom ParisTech, 23 avenue d'Italie, CS 51327,  75214 Paris CEDEX 13, France}

\address{$^3$Department of Physics and Astronomy, University of Exeter, Stocker Road,
Exeter EX4 4QL, United Kingdom.}

\date{\today}

\begin{abstract}
Measurement-based quantum computation (MBQC) and holonomic quantum computation (HQC) are two very different computational methods. The computation in MBQC is driven by adaptive measurements executed in a particular order on a large entangled state. In contrast in HQC the system starts in the ground subspace of a Hamiltonian which is slowly changed such that a transformation occurs within the subspace. Following the approach of Bacon and Flammia, we show that any measurement-based quantum computation on a graph state with \emph{gflow} can be converted into an adiabatically driven holonomic computation, which we call \emph{adiabatic graph-state quantum computation} (AGQC). We then investigate how properties of AGQC relate to the properties of MBQC, such as computational depth. We identify a trade-off that can be made between the number of adiabatic steps in AGQC and the norm of $\dot{H}$ as well as the degree of $H$, in analogy to the trade-off between the number of measurements and classical post-processing seen in MBQC. Finally the effects of performing AGQC with orderings that differ from standard MBQC are investigated.
\end{abstract}

\maketitle

\section{Introduction}

Quantum computation provides an advantage over classical computation in solving certain problems in shorter time. For example, known classical algorithms for factoring a number take exponential time in the number of inputs $N$, while a quantum algorithm polynomial in $N$ exists~\cite{Shor1997}. Three widely studied methods for implementing universal quantum computation are circuit-based quantum computation (see e.g.~\cite{NielsenChuang}), in which a series of unitary gates are applied to a number of input qubits; measurement-based quantum computation~\cite{Raussendorf2001} (MBQC), in which an entangled resource state is prepared and measurements are performed to drive a transformation on a portion of the state; and adiabatic quantum computation~\cite{Farhi} (AQC), in which the solution to a problem is encoded in the ground state of a Hamiltonian, and this ground state is reached adiabatically (i.e. slowly with respect to the minimum energy gap) starting from the ground state of some easily prepared Hamiltonian. These models have been shown to be equivalent to each other in the following sense; for a given computation, the number of gates required in the circuit model scales polynomially with the number of measurements on a graph state in MBQC~\cite{Raussendorf2001} (the size of the graph state) and polynomially with the inverse energy gap of the equivalent AQC computation~\cite{Aharonov2007}.

Bacon and Flammia proposed the direct translation of MBQC on a cluster state into an adiabatically driven HQC evolution~\cite{Bacon2010}, which they call \emph{adiabatic cluster-state quantum computation}. In this model, the initial adiabatic Hamiltonian is made up of the stabilisers of the cluster-state and the computation proceeds by replacing the discontinuous measurements of MBQC with continuous adiabatic transformations. While the evolution is adiabatic, Bacon and Flammia's model is not an example of an AQC. In AQC the computational task is to reach the unique ground state of a ‘problem’ Hamiltonian, in contrast, in Bacon and Flammia's model the ground state is degenerate, the subspace is known, but the evolution generates transformations within this degenerate subspace. This is an example of a holonomic quantum computation (HQC)~\cite{Zanardi1999,Pachos1999,Kult2006}. There are several other works combining ideas from MBQC and AQC. In ancilla-controlled adiabatic evolution~\cite{Wiebe2012,Kieferova2014,Hen2014}, computation is carried out by a combination of adiabatic passage and measurement. In adiabatic topological quantum computation~\cite{Cesare2014}, defects in a topological code are adiabatically deformed to perform logical operations. There are also examples of adiabatically driven computations on non-stabiliser states such as symmetry-protected states of matter~\cite{Williamson2014} and generalised cluster states~\cite{Brell2014}. 

Here we extend Bacon and Flammia's model to general graph states with a property called \emph{gflow}. This new \emph{adiabatic graph-state quantum computation} (AGQC) allows us to investigate how the properties of MBQC change when we replace non-deterministic measurements by deterministic adiabatic transformations. Beyond the application of the different models, it is of general interest to have methods of translating computations from one model to another so that intuition, understanding and techniques from one model can be applied to the others.

The random outcomes of the measurements in MBQC require a classical adaption of future measurements to achieve a deterministic outcome. This interplay allows for interesting trade-offs between classical and quantum time in MBQC and has given rise to new concepts such as blind quantum computation~\cite{Broadbent2009} and verified universal computation~\cite{Fitzsimons2012}. Furthermore it can be used to demonstrate a gap in quantum depth complexity between MBQC and the circuit model~\cite{Browne2011}. Here we are interested in how this trade-off manifests itself in AGQC. We find that it becomes a trade-off between the number of adiabatic steps and the degree of the initial Hamiltonian as well as the norm of the time derivative of the Hamiltonian. (The \emph{degree} is the number of sites that each summand in the Hamiltonian acts on non-trivially.)

Whether or not large degree operators act as a useful resource in this model is important from a fundamental point of view as well as a practical one. Our results suggest that large degree operators are not a useful computational resource in this model. We see that in order to decrease the number of adiabatic steps, the Hamiltonian degree needs to rise. However, this does not bring the benefit one may hope as the time needed for adiabatic passage through each step increases by the same amount so that the total time scales the same. Furthermore, under the assumption that simulating high degree operators shrinks the energy gap (which is the case for all known methods e.g.~\cite{Jordan2008,Bravyi2008}), implementation using fixed degree Hamiltonians will incur prohibitive time costs. This means that the optimal way of performing MBQC does not correspond to the optimal way to perform AGQC. These results are also interesting from the perspective of a subtlety that arises within the application of the adiabatic theorem. In our computation, the time taken for each step is governed not by the energy gap (which remains constant), but the norm of differential of the Hamiltonian.

Finally we are interested in how the time order associated to computation through MBQC appears in our model. One may expect that replacing random measurements with deterministic adiabatic substitutions will allow different ordering of the computational steps. In particular Clifford operations can be done in a single step in MBQC, so we expect this behaviour to manifest itself in AGQC. However we find that, in certain cases, the computational steps can be performed in a limited order. Surprisingly the most limited ordering occurs for certain Clifford operations.

The paper is structured as follows; in Section~\ref{sec:Background} we provide background on adiabatic quantum computation, measurement-based quantum computation, and adiabatic cluster-state quantum computation. In Section~\ref{sec:AQC} we generalise adiabatic cluster-state quantum computation to any graph state which has \emph{gflow}, and investigate what trade-off exists in this model in analogy to the trade-offs in MBQC. Finally we discuss the role of the ordering of measurements in adiabatic graph-state quantum computation in Section~\ref{sec:Order} and conclude in Section~\ref{sec:conc}.

\section{Background}\label{sec:Background}

\subsection{Adiabatic Holonomic Quantum Computation}\label{sec:AQC}

Consider a system with a time-varying Hamiltonian $H(t)$ and ground state $\ket{E_0(t)}$. If at $t=0$ we prepare a system in $\ket{E_0(0)}$, and change the Hamiltonian slowly enough, then at time $\tau$ we will finish in the state $\ket{E_0(\tau)}$ with high probability. In this case, `slow' means that the evolution satisfies the adiabatic criterion~\cite{Born1928}, which roughly says that the system will remain in the ground state provided that the evolution satisfies
\begin{eqnarray}\label{eqn:AdTheor1}
\frac{\bra{E_m(t)} \dot{H}(t) \ket{E_0(t)} }{ E_m(t) - E_0(t) } \ll 1
\end{eqnarray}
for all $m$ and $0 \leq t \leq \tau$. After such an adiabatic evolution, the final state is $e^{i \gamma_B(\tau)}e^{-i\int_0^\tau E(t')dt'} \ket{E_0(\tau)}$, where $-i\int_0^\tau E(t')dt'$ is the \emph{dynamical} phase and $i\gamma_B(\tau) = -\int_0^\tau \sprod{E_0(t')}{\dot{E}_0}dt' $ is the \emph{Berry} or \emph{Pancharatnam} phase~\cite{Berry1984}\cite{Pancharatnam1956}. The dynamical phase vanishes under cyclic evolutions where $H(0) = H(\tau)$ and can be removed by a local gauge transformation $\ket{\tilde{E}_0(t)} = e^{-i\int_0^\tau E(t')dt'} \ket{E_0(t)}$, however the Berry phase does not vanish under cyclical evolutions or local gauge transformations, and depends only on the path taken through parameter space during the evolution. The Berry phase has proved important when describing many phenomena in condensed matter systems, such as the anomalous quantum hall effect~\cite{Xiao2010}.

Berry phases can also be generalised to situations where $H(t)$ has ground space of $d$ degenerate energy levels, labelled $\ket{E_0^\alpha}$ where $1 \leq \alpha \leq d$. In this case, the same process will not in general lead to a global phase, but instead transforms the ground space by a rotation $U_{\alpha \beta}(t) = \mathcal{T} \exp \left( - \int_0^t \sprod{E_0^{\alpha}(t')}{\dot{E}_0^\beta(t')}dt' \right)$. This is called a \emph{holonomy}~\cite{Wilczek1984}, and for systems over which we have adequate control it is possible to use these holonomies to produce universal rotations on information encoded in the ground space, and so provides a way to perform quantum computation. This method is called \emph{holonomic quantum computation} (HQC)~\cite{Zanardi1999,Pachos1999}. It is also possible to perform HQC where $H(0) \neq H(\tau)$, this is known as {\em open-loop holonomic quantum computation}~\cite{Kult2006}.

Typically HQC is performed using an adiabatic evolution (although this is not a necessary condition~\cite{Aharanov1987}), but is a very different way of computation compared to the `standard' adiabatic quantum computation (AQC) protocol proposed by Farhi et. al.~\cite{Farhi2001}. The AQC protocol starts with a system prepared in the ground state of a simple initial Hamiltonian $H_0$, such as a uniform magnetic field, and then the Hamiltonian is slowly changed to a complicated `problem Hamiltonian' $H_p$ whose ground state encodes the problem to be solved (e.g. finding the ground state of an Ising spin glass is NP-hard~\cite{Barahona}). In HQC, it is not the  ground state itself that encodes the answer to the problem but the sequence of operations that have been performed within a degenerate subspace. The eigenstates of the Hamiltonian can be completely known at all times, as can the energy gap profile. AQC has a built in noise reduction mechanism since there is always an energy gap between the ground and states, however achieving fault tolerance in AQC is still an ongoing problem (see e.g.~\cite{Young2013}). HQC also has some gap protection (although due to the degeneracy in the ground state the protection is not the same as for AQC), and some protection from control errors (see e.g.~\cite{Carollo2003}). There are also known schemes to implement fault-tolerant HQC~\cite{Oreshkov2009}.

Since HQC involves degenerate ground spaces, the form of the adiabatic theorem shown in eqn. (\ref{eqn:AdTheor1}) is not appropriate since it is derived for singly degenerate states. For the purposes of this paper the following form of the adiabatic theorem is valid~\cite{Reichardt2004}; consider a linear interpolation between two Hamiltonians $H_0$ and $H_p$. The time dependent Hamiltonian of this transition is $H(t) = (1-\frac{t}{\tau})H_0 +\frac{t}{\tau} H_p$ so that the transition is finished at $t = \tau$. To simplify the notation, we introduce a parameter $s = \frac{t}{\tau}$, and we denote the eigenvalues of $H(s)$ as $E_n(s)$. Then if we start in the ground state of $H_0$, the final state will be $\varepsilon$ close in the $l_2$ norm to the ground subspace of $H_p$ provided the adiabatic run time $\tau$ satisfies
\begin{eqnarray}
\label{eqn:AdTheor2}
\tau \ge \max_{0 \le s \le 1}  \left( \frac{ c(\delta) \Vert\dot{H} \Vert^{1+\delta}}{\varepsilon \Delta^{2+\delta}} \right)
\end{eqnarray}
where $\Delta = \min_n|E_n(s) - E_0(s)|$, $0 < \delta \leq 1$, $\Vert M \Vert$ is the operator norm, defined as the largest absolute eigenvalue of $M$, and $c(\delta)$ is a parameter depending only on $\delta$. Although it is tempting to set $\delta \to 0$, this isn't possible without the adiabatic time diverging, since $\lim_{\delta _\to 0 } c(\delta) = \infty$~\cite{Reichardt2004}. So $\delta$ is taken as some fixed, small positive number.

\subsection{Measurement-based quantum computation}\label{sec:MBQC}

In MBQC an entangled resource state is measured sequentially and adaptively. We consider graph states as our entangled resource states (see e.g.~\cite{Gross2007} for other possible resources). To create this resource state, qubits are prepared in a $\ket{+} = (1/\sqrt{2})(\ket{0} + \ket{1})$ state, and controlled-phase ($CZ$) operations are performed between neighbouring qubits. To perform a computation on this resource state, single qubit measurements in bases $\{ \ket{+_{\theta_j}}\bra{+_{\theta_j}},\ket{-_{\theta_j}}\bra{-_{\theta_j}} \}$ are performed, where $\ket{\pm_{\theta}} := \frac{1}{\sqrt{2}} ( \ket{0} \pm e^{i\theta} \ket{1}),$ and where $\theta_j$ is the measurement angle for qubit $j$. The measurement will have a random outcome $\pm 1$, however by adapting future measurements on other qubits this randomness can be corrected for. The deterministic output of the computation is either encoded in the quantum state of the unmeasured qubits (that is, the quantum output), or in the classical measurement outcomes~\cite{Raussendorf2001,Anders2009}, with the former case giving a unitary evolution on the encoded information.

For example, consider a system of two qubits, A and B. Qubit A is prepared in a state $\ket{\phi} = \alpha \ket{0} + \beta \ket{1}$, whilst qubit B initially prepared in state $\ket{+}$. Performing  a $CZ$ gate between them entangles the input and results in the state $\ket{\psi_{AB}}= \alpha \ket{0}\ket{+} + \beta \ket{1} \ket{-}$. Note that by preparing A in state $\ket{\phi}$ instead of $\ket{+}$, we are able to encode information in the chain, so we call A the input to the chain, and since B is the system where the information will be at the end of the computation, we call this the output. We can rewrite the state $\ket{\psi_{AB}}$ as
\begin{eqnarray}
\ket{\psi_{AB}}&= \frac{1}{\sqrt{2}} \ket{+_{\theta}}( \alpha\ket{+} + e^{-i\theta} \beta  \ket{-}) + \frac{1}{\sqrt{2}} \ket{-_{\theta}}( \alpha\ket{+} -e^{-i\theta} \beta  \ket{-}) \nonumber\\
&= \frac{1}{\sqrt{2}} \ket{+_{\theta}} \tilde{H} U_z(\theta) \ket{\phi} +\frac{1}{\sqrt{2}} \ket{-_{\theta}} X \tilde{H} U_z(\theta) \ket{\phi},
\end{eqnarray}
where $U_z(\theta)$ is a rotation about the z-axis by angle $\theta$, and $\tilde{H}$ is a Hadamard gate. Now consider performing a $\ket{\pm_{\theta}}$ measurement on qubit A. If the outcome is $\ket{+_{\theta}}$, the state of qubit B is $\tilde{H} U_z(\theta) \ket{\phi}$, and if the outcome is $\ket{-_{\theta}}$ the state of qubit B is $X \tilde{H} U_z(\theta) \ket{\phi}$. If we apply a Pauli $X$ correction on qubit $B$ when the measurement outcome is $\ket{-_{\theta}}$, then both outcomes will be the same. In this way corrections allow for a deterministic implementation of the unitary operation $\tilde{H} U_z(\theta) \ket{\phi}$.

\begin{figure}[t]
\begin{center}
	\includegraphics[width=0.45\textwidth]{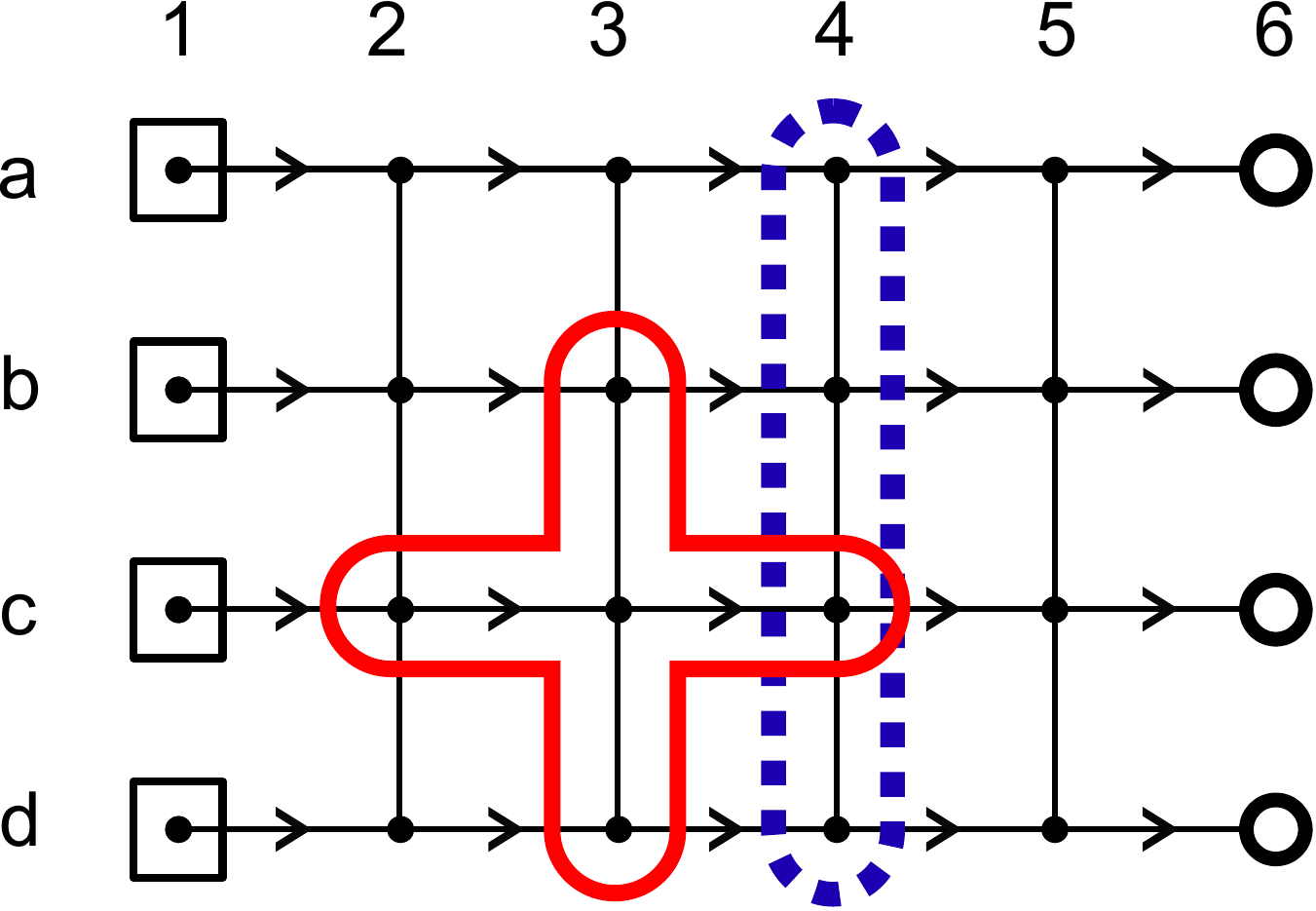}
	\caption{\label{fig:Example} An illustration of the definitions in Sec.~\ref{sec:MBQC}, applied to a cluster state with depth 5. Inputs are represented by vertices with squares, and outputs are vertices with hollow circles. Arrows indicate \emph{gflow} lines, the dotted circle indicates a single layer of qubits and the solid red cross contains all the vertices which contribute to $|g(v)|$ for vertex 2c (so $|g(2c)| = 5$).}
\end{center}
\end{figure}

More general computation in MBQC can be depicted using graphs. Qubits prepared in the $|+\rangle$ state are represented by vertices $V$ on a graph $G$, and the edges $E$ represent which pairs of qubits have been acted on by a $CZ$ gate. The state resulting from these operations is called a {\it graph state}, $\ket{G}$. The graph state of the cluster state, which is a universal resource for MBQC~\cite{Raussendorf2001}, is the two dimensional square lattice (see Fig.~\ref{fig:Example}). Graph states can also be defined using the stabiliser formalism~\cite{Hein06}, where it is defined as the state which satisfies the stabiliser eigenequations
\begin{eqnarray} \label{eqn:GSStabEQN}
K_v \ket{G} = \ket{G} \; \; \forall v \in V.
\end{eqnarray}
where a stabiliser generator is associated to each vertex $v$,
\begin{eqnarray} \label{eqn:GSStab}
K_{v}  &= X_{v} \prod_{w \sim v} Z_w.
\end{eqnarray}
Here $X_v$, $Y_v$, $Z_v$ are the Pauli matrices acting on site $v$, and the notation $v \sim w$ means that $v$ and $w$ are connected by an edge. The $K_v$ generate the \emph{stabiliser group} $S=\langle \{K_v\}_v \rangle$. The same group can be found by choosing different generators, for example $S=\langle \{K_\alpha K_v\}_v \rangle$ (for some arbitrary fixed vertex $\alpha$), and indeed the graph state is stabilised by the set of generators $\{K_\alpha K_v\}_v$ as well. This flexibility of choice will be useful later on.

As in the example above for a quantum input / quantum output computation, we can extend the definition to include input qubits, labelled $I$. In this case the stabiliser generators (\ref{eqn:GSStab}) are defined on non-inputs only. This is referred to as an \emph{open} graph state. During a computation all vertices are measured except the output vertices, labelled $O$. In this work we are concerned with computations for quantum inputs and outputs, so we will be using open graph states from now on.

In order for a measurement pattern on a graph state to be able to be correctable such that the output is the same regardless of the measurement outcomes, it is sufficient (although not in general necessary) for the graph to have \emph{Generalised flow} (\emph{gflow})~\cite{Danos2006,Browne2007}. Gflow is an incredibly useful tool in MBQC that has been used to study parallelism~\cite{Broadbent2009,Browne2011}, the translation between MBQC and the circuit model~\cite{Broadbent2009,Silva13} and the emergence of causal order in MBQC~\cite{Raussendorf2011}. \emph{Gflow} allocates a time ordering over the vertices on a graph state and a \emph{gflow function} $g(v)$ which tells us which vertices are affected by the measurement outcome of vertex $v$ and which qubits can be used to correct for this. It is defined for measurements in any of the three planes, $(X,Y)$, $(X,Z)$ or $(Y,Z)$, in a generalization of the example presented earlier. In this paper, we focus on measurements in the $(X,Y)$ plane, as results for measurements in other planes will follow in a similar fashion, although, for completeness we present the definition of \emph{gflow} for all planes. 

The notation $v< w$ is used to represent that vertex $v$ is measured after vertex $w$, and $v=w$ to indicate that $v$ and $w$ can be measured at the same time. We say that a set of vertices $U$ is oddly (evenly) connected to a vertex $v$ if there is an odd (even) number of edges connecting $U$ and $v$. The definition of \emph{gflow} is then:
\begin{defin}[Gflow]
Given an open graph state $G$ with inputs $I$, outputs $O$, edges $E$ and vertices $V$, we say it has \emph{gflow} if there exists a \emph{gflow function} $g$ and a time ordering $<$ over $V$ such that, for all $v \in V$ which are not outputs:
\begin{itemize}
\item[(G1)] All qubits $w$ in $g(v)$ are in the future of $v$, i.e. $v < w$ for all $w \in g(v)$.
\item[(G2)] if $w \leq v$, and $v \ne w$, then $w$ is evenly connected to all qubits in $g(v)$.
\item[(G3)]
\begin{itemize}
            \item $(X,Y)$ plane: $v\notin g(v)$, and $g(v)$ is oddly connected to $v$.
            \item $(X,Z)$ plane: $v\in g(v)$, and $g(v)$ is oddly connected to $v$.
            \item $(Y,Z)$ plane: $v\in g(v)$, and $g(v)$ is evenly connected to $v$.
\end{itemize}
\end{itemize}
\end{defin}
Given a \emph{gflow} $g$, the associated correction required after measurement of vertex $\nu$ with result $r_\nu$ is
\begin{equation}
\left( \prod\nolimits_{\mu \in g(\nu)}K_\mu \right)^{r_\nu}.
\end{equation}
In~\cite{Browne2006} it is shown that if an open graph has \emph{gflow} then it is possible to run an MBQC from input $I$ to output $O$. Indeed if we ask that the corrections work for all measurement angles on the planes, the existence of a gflow is necessary and sufficient. Finding out if a graph has \emph{gflow} can be done in polynomial time~\cite{Mhalla}.  Furthermore, the \emph{gflow} defines a valid measurement pattern, which may allow some qubits to be measured at the same time. Any qubits which can be measured simultaneously are said to be in the same layer of the computation. More formally;
 \begin{defin}[Layers]
 A layer of a computation is defined as any (non-output) qubits in a measurement pattern which can be measured at the same time.
\end{defin}
We denote the layers as $L_k$, and we use $L(v)$ to denote the layer that vertex $v$ is in. For example, for the \emph{gflow} defined on the graph in Fig.~\ref{fig:Example}, the layers are given by $L_k = \{ a_k,b_k,c_k,d_k,e_k\}$, for $k < 6$, and $L(a_k) = L(b_k) = L_k$ etc. We further denote $v\leq L_k$ for vertices in $L_k$ or earlier layers, similarly for $v\geq L_k$.
Using the concept of layers, we can define the depth for MBQC;
\begin{defin}[Depth]
The depth of an MBQC with \emph{gflow} is the number of rounds of measurements in the measurement pattern, or equivalently the number of layers in a measurement pattern.
\end{defin}
 In general this depth will be different depending on which \emph{gflow} we are using (there can be more than one - indeed we will see in Section~\ref{sec:Example} an example where many \emph{gflow}s can be realised). Since we can think of \emph{gflow} as a directed graph superimposed on an undirected graph, an equivalent and perhaps more intuitive definition is that the depth is the longest possible path along these directed edges.

The depth is effectively the time needed for the quantum part of the computation. To decrease this quantum time one is interested in pushing as many measurements together as possible to reduce the number of computational steps~\cite{Broadbent2009,Browne2011}. This is achieved by what is called the \emph{maximally delayed gflow}~\cite{Mhalla}. However, this is done at the expense of increasing the classical time needed to process the measurement results, which increases for larger \emph{gflow}. The techniques of \emph{gflow} thus give rise to a trade-off between the classical and quantum time for the computation~\cite{Broadbent2009,Browne2011}. As we will see in the example in Section~\ref{sec:Example} this trade-off can be great, so that for some cases all the time of the computation can be shifted to the classical processing except some constant quantum part. This trade-off is characterised in~\cite{Browne2011} in terms of circuits with fanout.

 We end this subsection with some further definitions which will be important in the theorems later on, and are illustrated in Fig.~\ref{fig:Example}.

\begin{defin}[$|g(v )|$]
The size $|g( v )|$ of a \emph{gflow} is the number of qubits for which the product of the stabilisers over $g( v )$, $\prod_{v\in g(v )}K_v$, is non-trivial.
\end{defin}
E.g. for the cluster-state, there is only one vertex in $g(v)$, and since each stabiliser generator acts on up to 5 qubits, the size of the \emph{gflow} is 5.

\begin{defin}[gflow lines]
A \emph{gflow line} is a directed edge from a vertex $v$ to an element of $g(v)$. The set of \emph{gflow} lines generates a directed graph over the set of vertices of the original graph (see for example Fig.~\ref{fig:Example}).
\end{defin}

\subsection{Adiabatic cluster-state quantum computation}\label{sec:ACSQC}

An adiabatically driven open-loop holonomic approach to MBQC was proposed in~\cite{Bacon2010}; adiabatic cluster-state quantum computation. The approach is similar to that used in~\cite{Oreshkov2009} for closed-loop HQC. First note that generating a cluster state is equivalent to initialising in the ground state of the Hamiltonian:
\begin{eqnarray} \label{eq:initialH}
H_0 \equiv - \gamma \sum_{v \in V} K_v,
\end{eqnarray}
where $\gamma$ parametrizes the strength of the interactions and $K_v$ are cluster-state stabiliser generators. Indeed this is also true if we replace the $\{K_v\}$ by any set of generators of the stabiliser group. This will be useful later when extending to general graph states with \emph{gflow}.

The computation protocol in~\cite{Bacon2010} proceeds first by preparing the system in the ground state of $H_0$, which could be done by preparing the system in the ground state of a uniform magnetic field and adiabatically changing to $H_0$. Then each stabiliser generator is replaced by a rotated Pauli-$X$ operators $X^{\theta_v} = \kebra{+_{\theta}}{+_{\theta}} - \kebra{-_{\theta}}{-_{\theta}} = e^{-i \theta_{v} Z_{v} / 2} X_v  e^{i \theta_{v} Z_{v} / 2} =e^{-i \theta_{v} Z_{v}} X_v$ (see Table~\ref{tab:Normal}), in analogy to the measurements in MBQC. This can be done one-by-one, or at the same time. If done one-by-one, the time for each individual replacement is independent of the computation, so that the total time scales with $N$. If all the replacements are done at the same time it becomes difficult to get analytical bounds on the time, however numerical studies suggest that the energy gap scales inversely with the number of qubits~\cite{Bacon}.

This model allows for implementation of single qubit rotations and CNOT gates, and so is universal for quantum computation~\cite{Barenco95}. For the purposes of this paper, we will often use twisted stabiliser generators $K_{n}^{\theta}$, defined as
\begin{eqnarray}
K_n^{\theta_{n}} = X^{\theta_{n}}_{n} \prod_{n\sim m} Z_m.
\end{eqnarray}
We use these to form an initial Hamiltonian $H_0=- \gamma \sum_{v \in \mathcal{V}} K_v^{\theta_{n}}$. Then instead of adiabatically replacing the stabiliser generators by $X^{\theta_n}_n$ operators, they are replaced by $X_n$ operators, and the resulting computation is the same. The advantage of using the twisted picture is that certain results become clearer compared to the untwisted version. Note that, although this choice of stabiliser generators does not really affect the analogous MBQC protocol, in adiabatic cluster-state quantum computation these stabiliser generators are different physical Hamiltonians which must get realised.

We now look at the single qubit rotation to illustrate the main ideas and introduce a few concepts. Details of doing a controlled-NOT gate are in the appendix, which completes the set of universal gates~\cite{Bacon2010}. To perform single qubit rotations in the adiabatic cluster-state model, consider a twisted one-dimensional cluster state. The stabiliser generators for such a state are
\begin{eqnarray}\label{Stab1D}
K_v^{\theta_{v}}  &=Z_{v-1} X^{\theta_{v}}_{v} Z_{v+1} \quad v = 2,...,N-1; \; \;K_N &=Z_{N-1} X^{\theta_{N}}_{N}.
\end{eqnarray}
To fit with the notation used in~\cite{Bacon2010}, we will use slightly different operators $\{ T_v \}$, where $T_v := K^{\theta_{v+1}}_{v+1}$. The initial Hamiltonian is then:
\begin{eqnarray}
H_0 \equiv - \gamma\sum_{v= 1}^{N-1} T_v.
\end{eqnarray}
$H_0$ has a doubly degenerate ground state, which we can use to encode a qubit. For example, we can define logical states $\ket{0}_L$ and $\ket{1}_L$ as
\begin{eqnarray}
\ket{0}_L := \prod_{v \sim w} CZ_{(v,w)} \ket{0}_1 \bigotimes_{n > 1} \ket{+_{\theta_n}}_n, \; \; \ket{1}_L := \prod_{v \sim w} CZ_{(v,w)} \ket{1}_1 \bigotimes_{n > 1} \ket{+_{\theta_n}}_n
\end{eqnarray}
and similarly $\ket{\pm}_L := \frac{1}{\sqrt{2}}(\ket{0}_L \pm \ket{1}_L)$ and $\ket{\pm i}_L := \frac{1}{\sqrt{2}}(\ket{0}_L \pm i\ket{1}_L)$. Preparing the ground state in a superposition $\alpha \ket{0}_L + \beta \ket{1}_L$ corresponds to attaching an input $\alpha| 0 \rangle + \beta|1\rangle$ to the cluster state in the MBQC picture.
The protocol proceeds by adiabatically replacing the $T_1$ with $X_1$ (see Table~\ref{tab:Normal}), so that the time-dependent Hamiltonian is
\begin{eqnarray} \label{Ht}
H(s) &=-  \gamma(1-s) T_1 - \gamma s  X _1 - \gamma \sum_{n =2}^{N-1} T_n
\end{eqnarray}
where $0 \le s \le 1$. After this process, the information is encoded in the new degenerate ground space of $H(1)$, and the information originally encoded in $\{ \ket{0_L}$, $\ket{1_L} \}$ has been transformed.

Rather than following how the information transforms by following how the ground state evolves, one can follow how the computation proceeds in terms of the logical operators $X_L$, $Y_L$, and $Z_L$, which for $H_0$ are
\begin{eqnarray}\label{eqn:LogOp}
X_L \equiv X_1 Z_2, \; Y_L \equiv Y_1 Z_2 , \; Z_L \equiv Z_1.
\end{eqnarray}
Following the transformations of these operators is equivalent to following how the ground states transform, and can be simpler to deal with. In this picture, encoding an arbitrary input state is equivalent to applying the field $H_{field} = \alpha X_L + \beta Y_L + \gamma Z_L$.

To see how the information is transformed by the adiabatic substitution in (\ref{Ht}), the logical operators are multiplied by stabilisers (since they act as identity) until they commute with $X_1$ (i.e.\ the logical operators are put in a form which is conserved during the adiabatic transformation). $X_L$ already commutes with $X_1$, but $Z_L$ doesn't so we multiply by $T_1$:
\begin{eqnarray}
Z_L \rightarrow Z_1 T_1 = X_2^{\theta_2} Z_3.
\end{eqnarray}

Then, after setting $X_1 \to \idop$ since we are in the $+1$ eigenstate of $X_1$, the logical operators become
\begin{eqnarray}\label{eqn:LogOp2}
X_L \rightarrow Z_2, \; Z_L \rightarrow X_2^{\theta_2} Z_3.
\end{eqnarray}
Now defining new logical operators in the same way as in eqn.~(\ref{eqn:LogOp}):
\begin{eqnarray}
X_L' \equiv X_2 Z_3, \;Z_L' \equiv Z_2
\end{eqnarray}
Expressing $X_L$ and $Z_L$ in terms of these new logical operators, we can see that the information has been moved one step along the chain and transformed by ${U}_{2}^{(L)} \tilde{H}^{(L)}$, where $\tilde{H}^{(L)}$ is a Hadamard operation acting in the logical subspace, and $U_{v}^{(L)} :=\exp[ -i\theta_{v} Z_L / 2]$ is a logical $Z$ rotation (c.f.~the example in Sec.~\ref{sec:MBQC}). At the next step we replace $T_2$ with $X_2$, and so on from left to right down the chain (see Table.~\ref{tab:Normal} for an illustration of this). Finally the information is encoded in the $N^{th}$ qubit, with the information transformed by the operation ${U}_{tot}^{(L)}$, where
\begin{eqnarray}\label{Transform}
U_{tot}^{(L)} = \tilde{H}^{(L)} \prod _{v = N-2}^2(U_{v}^{(L)} \tilde{H}^{(L)}),
\end{eqnarray}
and since $U_{v}^{(L)} = \exp[ -i\theta_{v} Z_L / 2]$, this will depend on the sequence of angles $\{\theta_v \}$ used (following~\cite{Bacon2010} we set $\theta_1 = 0$ for convenience).

Since we are doing adiabatic transformations, the speed of the computation is limited by the ratio of the energy gap and $\Vert \dot{H}(s) \Vert$. For the Hamiltonian in~(\ref{Ht}), this is given by (see~\ref{app:Stab})
\begin{eqnarray}\label{eqn:OneStep}
\tau \geq \tau_0 :=\frac{c(\delta)}{ \varepsilon 2^{1 + \delta/2} \, \, \gamma},
\end{eqnarray}
where $0 < \delta \leq 1$, and $\varepsilon$ is the error in the adiabatic evolution. For the remainder of this paper we will compare adiabatic evolution time to this time $\tau_0$, so this is our definition of one unit of time for the adiabatic computation.

\begin{table}[t]
\begin{eqnarray*}
\begin{array}{l | cccccc}
 & T_1 & T_2 & T_3 & \dots & T_{N-2} & T_{N-1}\\
& \downarrow &  &  &  & & \\
\mbox{Step 1} & X_1 & T_2 & T_3 & \dots & T_{N-2} & T_{N-1}\\
& & \downarrow &  &  &  & \\
\mbox{Step 2} & X_1 & X_2 & T_3 & \dots & T_{N-2} & T_{N-1}\\
& & & \downarrow  &  &  & \\
\vdots & & &   & \vdots &  &\\
& & & & & \downarrow  &  \\
\mbox{Step }  N- 2 & X_1 & X_2 & X_3 & \dots & X_{N-2} & T_{N-1}\\
& & & & & & \downarrow \\
\mbox{Step }  N- 1 & X_1 & X_2 & X_3 & \dots & X_{N-2} & X_{N-1}
\end{array}
\end{eqnarray*}
\caption{An illustration of the method to perform single qubit operations in adiabatic cluster-state quantum computation; arrows indicate an adiabatic transition from one operator to the other.}
\label{tab:Normal}
\end{table}

\section{Adiabatic graph-state quantum computation}\label{sec:AGQC}

In the previous section we reviewed the results of~\cite{Bacon2010,Bacon} where they show that universal quantum computation is possible using adiabatic substitutions, instead of measurements, on the cluster state. Here we generalise this method to other graph states using tools from MBQC and show that any MBQC measurement pattern on a graph state can be converted into an adiabatically driven adiabatic holonomic quantum computation of the form in~\cite{Bacon2010}, such that the same computation is performed. We call this \emph{adiabatic graph-state quantum computation} (AGQC). We will first consider doing step-by-step transitions, then explore how the trade-off between quantum and classical time in MBQC manifests itself in AGQC.

 \subsection{Translation of MBQC patterns with gflow to adiabatic computation}\label{sec:GflowAQC}

In AGQC, given an open graph state with \emph{gflow} $(g,<)$, and measurement angles $\{\theta_v\}$, we start in the ground state of the initial Hamiltonian
\begin{eqnarray}
H_0 =-\gamma \sum_v T_v,
\end{eqnarray}
where the $T_v$ are products of the twisted stabiliser generators,
\begin{eqnarray}
T_v := \prod_{w \in g(v)} K_w^{\theta_w},
\end{eqnarray}
$\forall v \notin O$. Hence the ground state corresponds to the twisted open graph state (an open graph state where the vertices are rotated according to the measurements). Preparing the system in the ground state of $H_0$ could be done by starting in the ground state of a magnetic field, and adiabatically evolving to $H_0$. This initial Hamiltonian can also be computed efficiently from the graph and \emph{gflow}, since each $T_v$ can be calculated in a time that scales as $O(\log (\mbox{max}_v |g(v)|))$ using the methods in~\cite{Aaronson2004}. For simplicity we take the number of inputs to be equal to the number of outputs $|I|=|O|$, but all statements and proofs can be easily extended to the cases $|O|>|I|$ ($|O|<|I|$ is not allowed as this would mean information is lost). The final Hamiltonian will be $H_f=-\gamma\sum_v X_i$, and the transition will be done in steps, as in the cluster-state case. Indeed if the graph is a cluster (2D lattice) our model reduces to adiabatic cluster-state quantum computation (and $T_v = K_{g(v)}$, as $g(v)$ contains only one element).

To perform the computation, the $T_v$ can be replaced one by one in an order that doesn't violate the \emph{gflow}, or those in the same layer can be replaced all at once. Replacing them one by one will take $N$ steps of time $\tau_0$. When replacing layer by layer, the adiabatic transition for the $k$th step is governed by the interpolation Hamiltonian
\begin{eqnarray}
H_{L_k}(s) = -\gamma \left(\sum_{v <  L_k}X_v +\sum_{v >L_k}T_v \right) - \gamma\left( \sum_{v \in  L_k} (1-s) T_v + s X_v  \right) \\
 = (1-s) H_{k-1} + s H_k,
\end{eqnarray}
where
\begin{equation}
 H_k= \gamma \left(-\sum_{v\leq L_k}X_v - \sum_{v>L_k}T_v \right).
\end{equation}

Note that $[X_u, T_v]_{u \neq v} = 0$ for any $u,v \in L_k$. Using this property and the analysis in~\ref{app:Stab}, we see that the time taken to perform the adiabatic evolution of Hamiltonian $H_{L_k}(s)$ scales with $\Omega (|U|^{1 + \delta} )$, where $0 < \delta \leq 1$. Interestingly the dependence of the time on $|U|$ does not come from the energy gap $\Delta$, which remains constant and independent of $|U|$, rather it comes from the norm $\Vert \dot{H} \Vert$, which scales as $|U|$. In this way we can replace all operators $T_v$ in the same layer simultaneously, but the adiabatic runtime for each layer scales as the size of each layer $|L_k|$. This point highlights a subtlety of the adiabatic theorem as applied here - the time is dominated by the norm $\Vert \dot{H} \Vert$, not the energy gap $\Delta$ as is more commonly the case.

Now consider what happens to the information when we replace all $T_v$ operators defined above with an $X_v$ operator, in the order given by \emph{gflow}. This can be seen by multiplying logical operators with stabilisers in such a way that the logical operators commute with all the adiabatic `measurements'~\cite{Bacon2010} (as illustrated for the one dimensional chain graph in Section~\ref{sec:ACSQC}). If there are any $Z_v$ or $Y_v$ operators which appear in a logical operator $\alpha$, the \emph{gflow} conditions guarantee that multiplying these terms by $T_v$ will either give the identity or an $X_v$ operator at vertex $v$, and the new logical operator will commute with any $X_w$ for $w \leq v$. This means that it is possible to update the logical operators $\alpha_L$, to $\tilde{\alpha}_L$ such that $[\tilde{\alpha}_L,X_v] = 0$ for all $v$ which are not outputs, and where $\alpha = X,Y,Z$. The output of the computation is encoded in these final logical operators, after setting $X_v \to \idop_v$ for all $v$ which are non-outputs.

To see that this performs the same computation as in MBQC, consider performing MBQC on a twisted graph state. If we start with the logical operators $X_L,Z_L$ of the MBQC resource state, and update these logical operators $\alpha_L$, to $\tilde{\alpha}_L$ such that $[\tilde{\alpha}_L,X_v] = 0$ for all $v$, then all measurements in the $X$ basis commute with these operators. Therefore if we start in the +1 eigenstate of the logical operators, after the measurements the final state will be the +1 eigenstate of $\tilde{\alpha}_L$ (after corrections have been applied). Indeed, the procedure outlined above for updating the logical operators is essentially the same process given by \emph{gflow} for tracing the logical operators in MBQC in the Heisenberg picture as in~\cite{Markham13}, thus the computation is clearly identical.

This is summarised in the following theorem.
\begin{theorem}
Any measurement based computation on an open graph state $\ket{G}$ of $N$ qubits which is the ground state of a Hamiltonian $H$ and which has \emph{gflow} g and depth $d$ can be efficiently converted into an adiabatically driven computation for which
\begin{itemize}
\item The adiabatic computation can be done in $d$ steps, where the energy gap for each step is the same, and $\Vert \dot{H} \Vert = |L_j|$ for the $j^{th}$ step. Thus the time to perform the $j^{th}$ step is $\Omega(|L_j|^{1 +
\delta})$, where $0<\delta \leq 1$.
\item The maximum degree of the initial Hamiltonian, $k_{\max}$, is equal to the maximum \emph{gflow} size : $k_{\max} = |g(v)|_{max}$,
\item The initial Hamiltonian can be computed efficiently.
\end{itemize}
 \end{theorem}
For example, for adiabatic cluster-state quantum computation on a rectangular graph qubits with $r$ rows and $d$ columns, $g(v) = v + r$, $T_v = K_{v+r}$, $|g(v) | \leq 5$ and there are $d$ layers, with each layer containing $r$ qubits, so the time for each layer scales as $\Omega(|r|^{1 +
\delta})$.

In the above theorem we have performed the computation step by step mimicking the measurement pattern in MBQC. We could also replace all of the $T_v$ operators at the same time regardless of which layers they are in, and the resulting computation would be the same, however the results in~\cite{Bacon} suggests that the energy gap for such an evolution would shrink polynomially in the depth. Since it is desirable to keep the gap as large as possible to provide protection against errors, and since it is hard to find (analytically or numerically) the energy gaps of systems other than simple 1-dimensional chains when doing a one step transition, we do not follow this approach here.

Note that, for all known universal graphs there is a \emph{gflow} for which $|g(v)|$ is bounded, so typically the degree will also be bounded. However this is not necessarily the case for all families of graphs; in some cases it can scale with the number of inputs (we will see an example of this in Section~\ref{sec:Example}). This raises an important question: Do we regard this increase in degree as a free resource, or is there a cost associated with it? The evidence to date would suggest that the latter is true, since typically in nature we only see 2-body interactions, with higher degree interactions resulting as a low energy approximation. We take this approach in our model, and assume that degree is bounded and such high degree Hamiltonians must be simulated. There are known methods for constructing such large degree operators from 2-local operators using \emph{perturbation gadgets}~\cite{Kempe2004,Oliveira2008,Bartlett2006,Jordan2008}. In particular we use the results from~\cite{Jordan2008}, that we can create k-local Hamiltonians $H_k$ using a perturbative Hamiltonian acting on $rk$ ancilla qubits and n computational qubits (r is the number of terms in Hamiltonian with degree $k$, which we consider as being fixed). The result is that the effective Hamiltonian, apart from some overall energy shift, is:
\begin{eqnarray}
\tilde{H}_{eff}(H^{gad}_{+} , 2^k, f(\lambda)) &= \frac{-k(-\lambda)^k}{(k-1)!} H_{k} \otimes P_{+}  + O(\lambda^{k+1}),
\end{eqnarray}
where $P_+$ is a projector on the space of $r$ ancilla qubits, projecting each one into the $\ket{+}$ state, and the perturbation converges provided that $\lambda < \frac{k-1}{4k}$.  

This energy gap decreases exponentially with $k$, therefore if we make the reasonable assumption that interactions in nature are limited 2 (or a finite number) of bodies, then if the degree is allowed to scale with $N$, this imposes a prohibitive cost in that the minimum energy gap of the system shrinks exponentially, and so therefore the adiabatic time grows exponentially. This is not a general result, since (as far as the authors are aware) there is no general theorem saying that the gap must shrink when approximating $k$-body Hamiltonians, but it is a intuitive result that we would expect to be true.

\subsection{Trade-offs in AGQC}
\label{sec:Example}

\begin{figure}[t]
   	\begin{center}
	\includegraphics[width=0.25\textwidth]{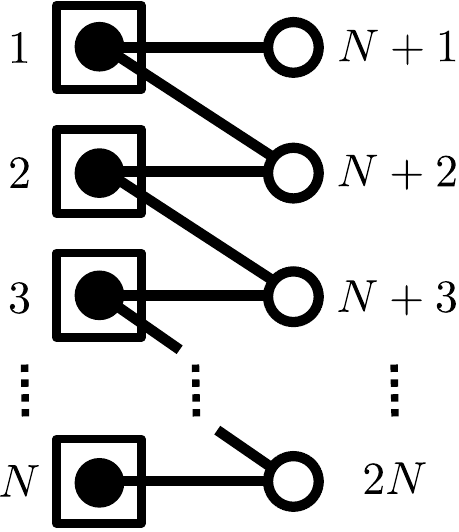}
	\caption{\label{fig:ZigZag} A graph for which many different \emph{gflows} are applicable. Vertices represent qubits and edges indicate that two qubits are entangled. Vertices inside boxes are input qubits, whilst hollow circles are outputs (which are not measured).}
	\end{center}
\end{figure}

In the previous subsection we saw how any MBQC computation with gflow can be mapped to a HQC. We now explore how the trade-off between quantum and classical time seen in MBQC~\cite{Browne2007} translates into this adiabatically driven model. First, as an illuminating example, consider the graph in Fig.~\ref{fig:ZigZag} which was presented in~\cite{Browne2007} and gives rise to a trade-off between classical and quantum times in MBQC. Many different \emph{gflows} can be defined on this graph, in particular a family of \emph{gflows} can be defined as
\begin{eqnarray}
g^r(v) = \left\{
\begin{array}{c c}
\{N+v,...,N+v+r -1 \}, & \mbox{if} \hspace{2mm}  v+r-1 \le N  \\
\{N+v,...,2N \},  & \mbox{if}  \hspace{2mm}  v+r-1 >N
\end{array} \right.
\end{eqnarray}
where $1 \le r \le N$. For a \emph{gflow} $g^r$, $r$ measurements can be performed simultaneously, interspersed by classical processing. Corrections on qubits will be of the form $X^{s_1 + s_2 + ... s_m}$ where the $s_m$ are binary variables accounting for the measurement outcomes. Thus the classical processing involves a binary sum of the results of these $r$ measurement outcomes, and so takes time $O(\log r)$~\cite{Furst84}. There is also classical processing required on the outputs, which involves the same number of terms to be added and so can also be done in time $O(\log r)$ (each output requires addition of $r$ binary variables or less, and these additions can be done in parallel). Since we can perform $r$ measurements simultaneously, the measurement depth $d^r$ is given by $d^r = \left\lceil \frac{N}{r} \right\rceil$, and the size of the \emph{gflow} is $|g^r(v)| \leq r+2$.

The two extreme cases are where $r=1$ or $r=N$. The former is just where each measurement is performed one-by-one ($d^1 =N$), with no addition of binary variables in between. The latter is where we can perform all measurements simultaneously ($d^N = 1$), but we must perform corrections on the outputs which take time $O(\log N)$. $g^N$ is called the \emph{maximally delayed flow} associated with this graph, whilst $g^1$ is the minimally delayed flow~\cite{Mhalla}. Since classical computation is typically a cheap resource, it is usually desirable to shift as much computation into classical processors as possible, and so in MBQC the optimal \emph{gflow} to choose would be the maximally delayed \emph{gflow}.

Following the conversion of these \emph{gflows} into an adiabatically driven computation, we can perform the computation in $d$ adiabatic steps, where the time to perform the $j^{th}$ layer scales as $\Omega(|L_j|^{1+ \delta} )$, and the maximum degree is given by the influencing volume. Thus $g^r$ is converted into an adiabatic computation which has $d^r$ steps, and each step takes $\Omega(r^{1+\delta})$ time, and the Hamiltonian degree $k = r+2$. In the most extreme case, $g^{N}$ is converted into an adiabatic computation which takes 1 step, but this step takes $\Omega(N^{1 + \delta})$ time to complete, and the Hamiltonian degree is proportional to $N$. In all cases the total time for computation is $\Omega(N^{1+\delta})$.

In this way the trade-off in quantum time ($d^r$) versus classical time ($\log (r)$) that is facilitated by the use of \emph{gflow} in MBQC, translates to a trade-off between the number of steps $(d^r)$ and the degree of the initial Hamiltonian ($r$), as well as the norm $\Vert \dot{H} \Vert$ ($\leq r+2$). However, even though the minimum energy gap is kept constant, the fact that the norm $\Vert \dot{H} \Vert$ is scales with $N$ means that there is no overall gain in time for the adiabatic computation. This result highlights the subtlety in the adiabatic theorem; although the minimum energy gap is the same for graphs of different flow, the size of $\Vert  \dot{H} \Vert$ changes and leads to a dependence of the adiabatic time on the number of elements in a \emph{gflow}. We also see that whereas in MBQC there is an advantage in using maximally delayed \emph{gflow}, there is no such advantage for the AGQC case. Indeed, given that the maximally delayed \emph{gflow} is accompanied by large (possibly unbounded) degree operators, and these Hamiltonians are likely to incur an exponentially decreasing energy gap, in AGQC it is better to use the minimally delayed \emph{gflow} instead.

In MBQC it is only possible to perform measurements in one step for certain computations. A natural question to ask is whether or not the trade-off seen above in the application of Theorem 1 to the zig-zag graph extends to more general computations, and can be extended to allow any computation to proceed in a constant time at the expense of high degree. If there was a way of implementing arbritrary computations instantly at the expense of increasing the degree only, then under reasonable assumptions this would imply an upper bound on the energy gap required for simulating such high degree operators. Such an upper bound would give an important fundamental limitation on the cost of large degree Hamiltonians, which does not currently exist.

To perform an AGQC on an arbitrary graph in one step, following the logic leading to Theorem 1, we must find a transformation of the stabilisers $T_v \to \tilde{T}_v$ such that all of the $\tilde{T}_v$ stabilisers  satisfy the commutation relations
\begin{eqnarray}
[\tilde{T}_v, X_w] = 0\; \mbox{for all} \; v \ne w \; \mbox{in} \; V \setminus O
\end{eqnarray}
Then all of the stabilisers $\tilde{T}_v$ can be replaced at the same time, provided the adiabatic runtime scales as $\Omega(|N|^{1+\delta})$. 

This transformation can be achieved by multiplying the $\tilde{T}_v$ stabilisers together. \emph{Gflow} always allows such a procedure, since whenever a stabiliser $T_w$ contains a $Z_v$ or $Y_v$ operator, these can be multiplied by $T_v$, giving identity or $X_v$, respectively, such that the resulting stabiliser commutes with $X_v$. Here we define a specific choice of such a procedure to make all stabilisers commute with all Pauli $X$ operators. To update stabiliser $T_v$, we start in layer $L_{n+1}$, where $L_n = L(v)$, and proceed as follows; (1) If $T_v$ contains a $Z_w$ or $Y_w$ term such that $w \in L_{n+1}$, multiply by $T_w$. (2) Proceed to the next layer, as determined by the time order. Iterate until the outputs are reached. (3) The final updated stabilisers are denoted $\tilde{T}_v$.

However, creating a Hamiltonian in this way is polynomially equivalent to simulating the computation by following the evolution of the logical operators (see e.g.~\cite{Markham13}), which we have already used in our discussion of making general measurement patterns adiabatic. Or to put it another way, this procedure is only efficient for certain classes of operations. For Clifford operations the procedure takes polynomial time; if we have $N$ qubits in total, we will have to perform one updating sweep per qubit, each sweep involves a search over at most $N$ stabilisers to see whether they commute or anti-commute, and the cost of testing if a stabiliser commutes or not will be $O(N)$ since each stabiliser contains at most $N$ terms, so the overall procedure takes $O(N^2)$ steps. For general angles the procedure takes an exponential amount of time, since at every step we replace a $Z_v$ operator with a $Z_vT_v = e^{-i \theta_{v+1} Z_{v+1}} X_{v+1} Z_{v+2}$ term, so every update converts a $Z$ term into 3 new terms after expanding the exponent. If we start off with a stabiliser containing $n$ $Z$ operators, then after $r$ sweeps we will have $O(3^r n)$ operators to search through.

In addition, during this update procedure, we multiply every $Z_w$ operator by $T_v = \prod_{w\in g(v)} K_{w}$. So following this procedure, each $\tilde{T}_v$ operator will have Pauli $X$'s in positions $g(w)$ for every $Z_w$ that has been have corrected for. The only parts that will contribute to the degree of $\tilde{T}_v$ are situated at vertices which can be arrived at by following a path on which a non-gflow line is preceded and followed by a gflow line. Following the discussion in Sec.~\ref{sec:GflowAQC} we expect the simulation of these Hamiltonians to be very prohibitive. 

As noted in Sec.~\ref{sec:GflowAQC}, an alternative way to perform all operations in one step is to just replace all stabilisers at once, without changing the form of the stabilisers, and we would expect the energy gap to shrink polynomially in $N$ for AGQC~\cite{Bacon}. This alludes to another complementary trade-off between the energy gap and the number of steps in the computation.

In summary, we see that in AGQC it is always possible to perform the computation in one step, at the expense of increasing $\Vert \dot{H} \Vert$ and the degree of operators in the initial Hamiltonian. This means that there is no decrease in the overall computation time. In addition, the process of finding the initial Hamiltonian is closely linked to the classical simulation of the computation itself, and thus can also only be efficiently calculated for simple cases such as Clifford computations. We see that reducing the number of steps in the computation has no advantage in overall computation time. Furthermore, given the exponentially decreasing energy gap for the known methods of simulating large-degree operators, this suggests that the optimal way to perform AGQC is to keep the degree as low as possible, which in general means as many steps as possible. In the language of MBQC, this corresponds to using the minimally-delayed gflow for the computation.

\section{Reordering the computation} \label{sec:Order}

In general, performing measurements in the wrong order in MBQC results in random outcomes to the computation. In some cases however the order of measurement does not matter, such as when all qubits are measured in the Pauli basis. A natural question to ask is whether or not this property also applies to AGQC. In this section we will go back to the original formulation of adiabatic cluster-state quantum computation, and look at what order the adiabatic substitutions can be performed in. Firstly we will consider the most obvious case, where the stabilisers are exactly the same as in~\cite{Bacon2010}, one stabiliser is replaced by one Pauli operator at each step. Then we consider a less constrained method, and discuss how these two approaches lead to different behaviour.

\subsection{Re-ordering the adiabatic computation with fixed number of terms in Hamiltonian}\label{sec:Order2}

We start with an $N$-qubit 1D chain with $N-1$ stabiliser generators (i.e. one qubit is encoded in the chain), and we consider replacing one stabiliser generator by one Pauli operator in a different order to the corresponding measurement pattern in MBQC. As a physical motivation we may imagine that we have some experimental apparatus which is limited to only applying the $T$ and $X$ operators, and which can turn them on in any combination. We would like to find orders of replacements which keep the system in the 2-dimensional logical subspace, and we might expect from MBQC that changing the order will in some way disrupt the computation. As an example, consider a 4-qubit chain with all angles set to zero (i.e. with untwisted stabiliser generators). The system is initialised in the ground state of the Hamiltonian
\begin{eqnarray}
H_0 = -T_1 - T_2 - T_3.
\end{eqnarray}
Clearly it is possible to replace $T_2 \to X_2$ out of order, since $[T_1,X_2] = 0$. However, if $T_3$ is replaced with $X_3$ out-of-order, this leads to the time-dependent Hamiltonian
\begin{eqnarray}\label{eqn:ReOrder1}
H(s) = -T_1 - T_2 - (1-s)T_3 - sX_3,\; \mbox{ with } 0\le s \le 1.
\end{eqnarray}
At $s=1$, this Hamiltonian has a ground state degeneracy of 4, i.e.\ the degeneracy doubles. However, there is a constraint in that the operator $T_1T_3$ commutes with $H(s)$ for all $s$ and so the eigenstate of $T_1T_3$ is conserved throughout the evolution. Therefore, since the system starts in the $+1$ eigenspace of $T_1T_3$, it will also end in the $+1$ eigenspace of $T_1T_3$. So although the gap closes, the transitions between these degenerate eigenstates are forbidden and so the logical subspace is preserved.

Now consider replacing $T_1$ in (\ref{eqn:ReOrder1}) with $X_1$. Throughout this evolution, $T_1T_3$ no longer commutes with $H(s)$, and so unless there is a stabiliser generator that can be multiplied with $T_1T_3$ to make it commute with $H(s)$, the evolution is no longer constrained to the $+1$ eigenspace of $T_1T_3$. It is easy to check that there are no remaining stabiliser generators which satisfy this property, since this requires a generator with a $Z_1$ term (and $T_1$ cannot be used since it doesn't commute with $X_3$ or $X_1$). Since at the very start of the $T_1 \to X_1$ evolution, the energy gap is zero and the subspace is no longer preserved, this means the information now leaks out of the subspace unless the evolution time $\tau \to \infty$. Thus the computation fails at the final step for this ordering. Note that this happens whether or not $T_2$ is replaced by $X_2$; the key part was the fact that there were no stabiliser generators left to multiply $T_1 T_3$ with to make it commute with $H(s)$.

Extending this to larger chains, whenever $T_n$ is replaced with $X_n$ such that $n > 2$, the evolution is still constrained to the $+1$ eigenspace of $T_{n-2}T_n$. If this `hidden' stabiliser anticommutes with the next Pauli replacement $X_m$, then provided we can multiply by $T_{m-2}$ the subspace is still preserved. But since at some point we need to replace $T_{1} \to X_1$ or $T_2 \to X_2$, and there are no other stabiliser generators to multiply with to make sure that the `hidden' stabiliser still commutes with $H(s)$, the computation fails as there will be zero energy gap and a non-zero matrix element to leak out of the 2-dimensional logical subspace.

Similar behaviour can be seen when the angles $\{ \theta_v \}$ are all odd multiples of $\pi/2$ (corresponding to $Y$ measurements). Again, using untwisted stabiliser generators, consider a Hamiltonian on 3 sites, $H = T_1 + T_2$. After replacing $T_2 \to Y_2$, the system is still in a 2-dimensional subspace since $[T_1T_2,H(s)]=0$ and so the subspace where $T_1T_2$ has eigenvalue $+1$ is preserved. However, after replacing $T_{1} \to Y_1$, there are no stabiliser generators which can be chosen, and so the computation fails at the last step. This argument can similarly be extended to larger chains, and just like in the above case we will find that once we reach the boundary the computation time has to go to infinity to avoid leakage.

Similar behaviour is also seen for the CNOT gate proposed in~\cite{Bacon2010} and discussed in~\ref{sec:CNOT}. Consider two rows of qubits, labelled $a$ and $b$, with 3 columns numbered from left to right. The initial Hamiltonian is:
\begin{eqnarray}
&H_{CNOT} = -T_{a1} -T_{a2} - T_{b1} - T_{b2} \nonumber\\
&-Z_{a1}X_{a2}Z_{a3} Z_{b2} - Z_{a2} Z_{b1} X_{b2} Z_{b3} - Z_{a2} X_{a3} - Z_{b2 }X_{b3}.
\end{eqnarray}
Although $T_{a2} \to X_{a2}$ can be replaced out of order without leaking out of the 4-dimensional subspace (since the operator $T_{b1}T_{a2}$ is conserved), then when $T_{a1} \to X_{a1}$ is replaced afterwards, the computation fails. Thus we have seen that, with this approach, the only replacements in 1 dimension which can preserve the subspace are replacements with $X_{n}$ with $n \le 2$, and replacements by $Y_n$ with $n \le 1$, and no out-of-order measurements are possible for the CNOT gate.

What about more general angles? Since both $X_n$ and $Y_n$ measurements fail when $n > 2$, and any general measurement is a superposition of $X$ and $Y$ measurements, we might expect that any general measurement fails for $n >2$. So we try replacements starting on the second site for general angles, i.e. replacements of the form
\begin{eqnarray}
-\sum_{n=1}^{N-1} T_n \rightarrow - T_1 - X^{\theta_2}_2 - \sum_{n = 3}^{N-1} T_n
\end{eqnarray}
The energy gap $\Delta_1$ for such a process is given by
\begin{eqnarray}
\Delta_1 (\theta_{2},s) &= \sqrt{ 2(1 - s + s^2) + \Gamma(\theta_{2},s)} -  \sqrt{ 2(1 - s + s^2) - \Gamma(\theta_{2},s)}.
\end{eqnarray}
where $\Gamma (\theta,s) \equiv \sqrt{ 2s^2 \cos{2\theta_{2}} + (4 - 8s + 6s^2)}$. For a given $\theta_{2}$, this reaches a minimum at $s = (1- \frac{1}{2}\cos \theta_{2})$. Thus for all $\theta_{2}$, the minimum energy gap $\Delta_1^{min}$ is given by:
\begin{eqnarray}
\Delta_1^{min}(\theta_{2})= \Delta_1 \left(\theta_{2} ,1- \frac{1}{2}  \cos \theta_{2}\right).
\end{eqnarray}
Over all possible $\theta_{2}$ values, $\Delta_1^{min}(\theta_{2})$ is largest for $\theta_{2} = l\pi, s = 1/2$, and goes to zero when $\theta_{2}= (2l+1)\pi/2, s=1$, where $l$ is an integer.

So we see that, for angles $\theta_2$ such that the minimum energy gap remains non-zero, it is possible to perform the computation out-of-order. While the information remains in a protected subspace, it is also necessary to check if the information is transformed in the expected way. To see this, we start with the logical operators $X_L \equiv X_1 Z_2, \; Z_L \equiv Z_1 $ (excluding $Y_L$ since $Y_L = iZ_L X_L$). Using the method in~\cite{Bacon2010}, these operators are multiplied by stabilisers to make them commute with the `measurements'. The appropriate transformation for the above case would be
\begin{eqnarray}
&X_L \to X_L T_2 = X_1 X_3^{\theta_3}Z_4 \nonumber\\
&Z_L \to Z_LT_1 =  e^{-i\theta_2 Z_2} X_2 Z_3 \to  e^{-i\theta_2 Z_2 T_2} X_2 Z_3 =  e^{-i\theta_2 X_3^{\theta_3}Z_4} X_2 Z_3.
\end{eqnarray}
This is exactly the same transformation we would make in the normal computation, except two steps have been performed at once. Both of these logical operators are in a form which commute with $X_1$ and $X_2$, so they commute with the time-dependent Hamiltonian, and so, for example, if the system starts in the $+1$ eigenstate of $X_L$, it will end up in the $+1$ eigenstate of $X_3^{\theta_3}Z_4$.

In summary, we have seen that if we simply replace $T_n$ stabiliser generators with $X_n$ operators one-by-one on a 1D chain, we are extremely restricted in what we can do, and reordering is only possible using measurement angles which are not odd multiples of $\frac{\pi}{2}$. In particular, in a 1D chain, we can only start by replacing the stabiliser at the first site or the second site. The information stored in the chain is transformed in the same way in both cases, however the latter only works for $\theta_2 \ne (2n+1) \frac{\pi}{2}$, and the energy gap depends on $\theta_2$ so the speed of the adiabatic substitution must vary.

\subsection{Re-ordering without a fixed number of terms in Hamiltonian}

We have seen that the initial approach to re-ordering the operations works only for a limited case. This is perhaps expected, as the way we have performed the out-of-order operations so far is not a proper reflection of what happens in MBQC. For instance, in MBQC measuring a qubit destroys any entanglement on edges connected to that qubit. This is clearly not true above, since we end up with Hamiltonians containing terms such as $T_1 + T_2 + X_3$. So a more natural way to perform the out-of-order measurements would be to remove all entanglement to measured qubits, or just remove all anticommuting terms entirely~\cite{Fitzsimons}. Take the chain considered above:
\begin{eqnarray}
H_0 = -T_1 - T_2 - T_3.
\end{eqnarray}
Now if $T_3$ is replaced by $X_3$, any other operators which anticommute with $X_3$ are also removed, i.e. the final Hamiltonian is
\begin{eqnarray}\label{eqn:ReOrder2}
H_1 = - T_2 - X_3.
\end{eqnarray}
Notice that the operator $T_1 T_3$ is still conserved, as in the previous subsection. Instead of replacing $T_1$ with $X_1$, $-X_1$ can be added to the Hamiltonian, and the system is still constrained to be within a 2-dimensional subspace. This would also work if we had instead just removed all entanglement to site $3$ (except $T_2$ can be left untouched since it commutes with $X_3$), i.e.
 \begin{eqnarray}\label{eqn:ReOrder3}
 H_1 = -Z_1 X_2 - T_2 - X_3.
 \end{eqnarray}
The energy gap will still be the same when we introduce $X_1$ in equation (\ref{eqn:ReOrder2}) to when we replace $Z_1X_2\to X_1$ in equation (\ref{eqn:ReOrder3}), since the Hamiltonian is of the form $(1-s)Z_1 \otimes A + sX_1$ and so has energy gap $2\eta = 2\sqrt{1 - 2s +2s^2}$ (\ref{app:Stab}).

Similar results hold for $Y$ measurements or the CNOT gate, so that Clifford operations can be performed in any order. Note that if more than one operator is replaced at the same time, the number of anticommuting terms increases, and based on the numerical studies in~\cite{Bacon}, if the number of anticommuting terms scales with $n$ we would expect the energy gap to be polynomial in $1/n$.

In summary, we have seen that Clifford operations can be done in any order, provided that either the stabiliser generators are modifed so that the entanglement to `measured' sites is destroyed, or any stabiliser generators which anticommute with the measurement operator are removed entirely. This is in contrast to the method in the previous subsection, in which the stabiliser generators did not reflect what happens in MBQC, and in which the re-ordering is limited to performing alternating measurements from left to right. In both of the approaches considered, Clifford operations cannot be performed all in one step, since we would expect the energy gap to decrease polynomially in the number of stabiliser generators we replace.

\section{Conclusions} \label{sec:conc}

We have shown that any measurement pattern with \emph{gflow} can be converted into an adiabatically driven holonomic quantum computation, such that the number of adiabatic steps is equal to the depth in MBQC, and each step takes time proportional to the size of each layer. This opens up the possibility of future results about efficiency and trade-offs in MBQC being used in AGQC. For example, there is still little understood about how to view the efficient simulatability of Clifford gates from the perspective of \emph{gflow}. The framework developed here offers a natural route to translate future possible results in this area, which may have implications for the efficiency of AGQC as well as the simulation of many-body Hamiltonians. In addition, since fault-tolerant schemes for HQC on stabiliser codes exist~\cite{Oreshkov2009}, we would expect it to be possible to extend this to AGQC. Beyond computation itself, the inherent interplay between classical and quantum processing in MBQC has led to the cryptographic protocols of blind quantum computation~\cite{Broadbent2009} and verified universal computation~\cite{Fitzsimons2012}. Our translation captures much of this interplay by using gflow as a main tool, so one may hope that it can help translate these protocols and ideas across to HQC.

We have also found that, in analogy to the trade-off between quantum and classical time in MBQC, there is a trade-off between the number of adiabatic steps taken and the size of $\Vert \dot{H} \Vert$, together with the degree of the initial Hamiltonian. One interesting point is that the trade-off is only between the number of steps and $\Vert \dot{H} \Vert$, but does not involve the energy gap, which highlights the subtleties of using the adiabatic theorem. It is perhaps surprising that, even if we could simulate high-degree operators without an exponentially shrinking energy gap, the overall adiabatic time would still scale with $N$. We might have expected that the free availability of large degree operators would allow some computational speed up, but the fact that this isn't the case suggests that there may not be any reason \emph{in principle} why it should be prohibitive to simulate large degree operators. It is also interesting that the increase in degree does not appear to provide any computational advantage in this model. Since known methods of simulating high-degree operators result in an exponentially small energy gap in AGQC it is optimal to use the minimally delayed flow, in contrast to MBQC where the maximally delayed flow is preferable.

We considered the influence of adiabatic measurements in a different order from the corresponding MBQC pattern. When stabilisers are replaced in a different order, the computation fails with the exception of a few, special cases. When however the operations are performed in such a way that the terms that anticommute with the measurement are all adiabatically removed, then all Clifford operations can be performed in any order.

Finally we stress that these results do not cover all possible methods of performing MBQC, as there are some graph states without \emph{gflow} which still yield deterministic computations~\cite{Browne2007}. For both Theorems we require correcting sets to be known which allow us choose stabilisers $\{T_v\}$ such that $\{T_v,X_v\} = 0$, and $[T_v,X_w] = 0$ for all $v \ne w$. For extensions of \emph{gflow} where such correcting sets are known our arguments can be carried forward simply, for example for Pauli Flow~\cite{Browne2007}. However, more generally, the lack of characterisation of possible correcting sets means it is not easy to guarantee these conditions are met so our procedure may not work. There are also more resource states for MBQC such as those investigated in~\cite{Gross2007}, which could have interesting properties.

\ack

The authors would like to thank Dan Browne, Elham Kashefi, Joe Fitzsimons and Terry Rudolph for illuminating discussions. BA is funded by the EPRSC. DM is funded by the FREQUENCY (ANR-09-BLAN-0410), HIPERCOM (2011-CHRI-006) projects, and by the Ville de Paris Emergences program, project CiQWii. JA acknowledges support from the Royal Society in the form of a Dorothy Hodgkin Fellowship (DH080235).

\section*{References}
\bibliography{AdiabaticBib}
\bibliographystyle{iopart-num}

\appendix
\section{Adiabatic CNOT gate}\label{sec:CNOT}

\begin{figure}[h]
\centering
\includegraphics[scale = 0.7]{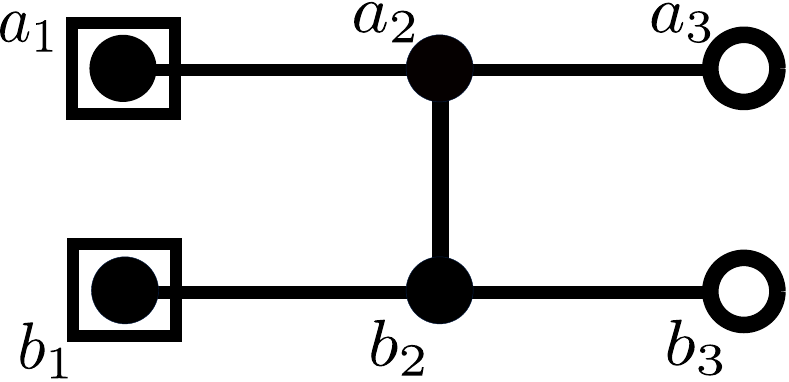}
\caption{An illustration of the graph used to perform a CNOT gate in adiabatic cluster-state quantum computation.}
\label{fig:CNOT}
\end{figure}

To perform universal quantum computations an entangling gate on the encoded information is needed, such as a controlled-NOT gate. Consider two rows of qubits, labelled $a$ and $b$, with 3 columns numbered from left to right (see Fig.~\ref{fig:CNOT}). The protocol starts with an initial Hamiltonian in which all angles $\{ \theta_n \}$ are 0, and which the inputs are on qubits $a_1$ and $b_1$, and the outputs are on qubits $a_3$ and $b_3$.:
\begin{eqnarray}
H_{CNOT} &= -Z_{a1}X_{a2}Z_{a3} Z_{b2} - Z_{a2} Z_{b1} X_{b2} Z_{b3} - Z_{a2} X_{a3} - Z_{b2 }X_{b3}.
\end{eqnarray}
Adiabatically replacing all of these stabilisers with $X$ operators results in a CNOT gate acting on the encoded information~\cite{Bacon2010}. Each replacement of a single stabiliser by a local Pauli operator still has the same adiabatic time as in the previous section, i.e.\ the adiabatic evolution time takes the form $\tau_0$, provided these substitutions are done progressing from left to right. A controlled-$Z$ rotation can also easily be achieved using an adiabatic scheme based on the gates in~\cite{Browne2007}, in which only 3 qubits are required to perform a gate on two qubits.

\section{}\label{app:Stab}

In this section we will prove the following lemma

\begin{lemme}
Consider two Hamiltonians acting on a graph with vertices $V$ and \emph{gflow} $(g,<)$;
\begin{eqnarray}
H =- \gamma \left(\sum_{u<  U} X_u + \sum_{v \geq U} T_v \right), \hspace{1cm}H' = - \gamma \left(\sum_{u \leq  U} X_u + \sum_{v > U} T_v \right)
\end{eqnarray}
where $T_v := \prod_{w \in g(v)} K_w^{\theta_w}$. At time $s=0$ we prepare a state in the ground subspace of $H$, and adiabatically change the Hamiltonian to $H'$ using an interpolation of the form $H(s) = (1-s)H + sH'$. Then provided $[T_u,X_v]_{v \ne u} = 0$ for all $u,v \in U$, it is possible to finish in the ground subspace of $H'$ with high probability provided that the adiabatic time $\tau$ can scale as $\tau = \Omega( |U|^{1 + \delta} )$ where $0 \le \delta \leq 1$~\cite{Reichardt2004}.
\end{lemme}

To prove this, the spectrum of $H(s)$ and the spectrum of $\dot{H}(s)$ must be derived and inserted into the expression for the adiabatic time in equation (\ref{eqn:AdTheor2}). The time dependent Hamiltonian in this case is
\begin{eqnarray}\label{eq:AppHam}
H(s) &=  -\gamma \sum_{u \in U} [(1-s)T_u +  sX_u]  -\gamma  \sum_{u < U} X_u -\gamma  \sum_{u > U} T_u 
\end{eqnarray}
First note that $X_v$ commutes with all stabilisers $T_w$ with $w \geq v$, $w\neq v$, and anticommutes with $T_v$. This follows from the conditions of \emph{gflow}, since a product of stabiliser generators $\prod_{y \in g(w)} K_y$ has an even number of connections to vertices $v \leq w$ (which means an even number of $Z$ operators, which cancel to give identity), whilst $T_v$ has an an odd number of connections to vertex $v$ (which means there is one $Z_v$ term $T_v$). This is true whether or not we are considering twisted stabilisers. Therefore we have $\{T_v,X_v\} = 0$, and $[T_w,X_v] = 0$ for all $w \geq v$, which means that the terms in the second two sums commute with each other and with the first sum. Then since $[T_u,X_v]_{v \ne u} = 0$ for all $u,v \in U$, each of the summands in the left hand sum commute with each other. We therefore know that the eigenvalues of $H(s)$ will be formed from combinations of the eigenvalues of the $\gamma [(1-s) T_u +  sX_u]$ terms, plus multiples of $\pm 1$ from the remaining terms. To be exact, since the terms in $H(s)$ are commuting normal operators with non-overlapping spectral projections, following the analysis in~\cite{Nielsen2005} the eigenvalues of $H(s)$ will be all possible combinations of the eigenvalues of the individual commuting terms. The eigenvalues of the individual $[(1-s) T_u +  s X_u]$ terms are $\pm \sqrt{ (1-s)^2 +s^2 } := \pm \eta$, so the eigenvalues of $H(s)$ are $-|U|\gamma\eta,-(|U|-2)\gamma\eta,...,(|U|-2)\gamma\eta,|U|\gamma\eta$ plus some integer multiples of $\pm 1$ from the remaining $X_n$, $T_n$ terms. The energy gap between the ground state and first excited state is therefore $2 \eta$. This is minimal at $s=\frac{1}{2}$, at which point the gap is $\sqrt{2}\gamma$.

The time-derivative of the Hamiltonian is
\begin{eqnarray}
\dot{H}(s)=\gamma\sum_{u \in U} ( T_u - X_u).
\end{eqnarray}
Using the same reasoning as above, the eigenvalues of this derivative are $\{ -|U|\gamma,-(|U|-2)\gamma,...,(|U|-2)\gamma,|U|\gamma \}$, so the norm of $\dot{H}(s)$ is just $|U|\gamma$. Inserting the minimal energy gap and $\Vert \dot{H}(s) \Vert$ into equation (\ref{eqn:AdTheor2}), we then find that a system prepared in the ground state of $H$ which is adiabatically changed to $H'$ using a linear interpolating function will be $\varepsilon$-close in the $l_2$ norm to the ground subspace of $H'$ provided that the adiabatic time $\tau$ scales as
\begin{eqnarray}
\tau \ge  \left(\frac{c(\delta) |U|^{1 + \delta} }{\varepsilon \, 2^{1 + \delta/2} \, \gamma} \right) = \Omega(|U|^{1 + \delta})
\end{eqnarray}

\end{document}